\documentclass[twocolumn, times]{aastex631}

\usepackage{amsmath}
\usepackage{amssymb}
\usepackage{bm}
\usepackage{upgreek}
\usepackage{graphicx}
\usepackage{float}
\usepackage{natbib}

\newcommand{\abs}[1]{\lvert #1 \rvert}

\newcommand{\bigparen}[1]{\left ( #1 \right )}
\newcommand{\bigcurly}[1]{\left \{ #1 \right \}}
\newcommand{\bigbracket}[1]{\left [ #1 \right ]}
\newcommand{\DMunits}{\,pc\,cm$^{-3}$}
\def\soft#1{\texttt{#1}}

\shorttitle{Detecting Fast Radio Bursts with spectral structure}
\shortauthors{Kumar, Zackay, \& Law}

\begin{document}
\title{Detecting Fast Radio Bursts with Spectral Structure using the Continuous Forward Algorithm}

\correspondingauthor{Pravir Kumar}
\email{pravir.kumar@weizmann.ac.il}

\author[0000-0003-1913-3092]{Pravir Kumar}
\affiliation{Department of Particle Physics and Astrophysics, Weizmann Institute of Science, 76100 Rehovot, Israel}

\author[0000-0001-5162-9501]{Barak Zackay}
\affiliation{Department of Particle Physics and Astrophysics, Weizmann Institute of Science, 76100 Rehovot, Israel}
\affiliation{Benoziyo Center for Astrophysics, Weizmann Institute of Science, 76100 Rehovot, Israel}

\author[0000-0002-4119-9963]{Casey J.~Law}
\affiliation{California Institute of Technology and Owens Valley Radio Observatory, California, USA}

\begin{abstract}
Detecting fast radio bursts (FRBs) with frequency-dependent intensity remains a challenge, as existing search algorithms do not account for the spectral shape, potentially leading to non-detections. We propose a novel detection statistic, which we call the Kalman detector, that improves the sensitivity of FRB signal detection by incorporating spectral shape information. The detection statistic is based on an optimal matched filter, marginalizing over all possible intensity functions, weighted by a random walk probability distribution, considering some decorrelation bandwidth. Our analysis of previously detected FRBs demonstrates that the Kalman score provides a comparable yet independent source of information for bursts with significant spectral structure, and the sensitivity improvement is of the order 0\%--200\% with a median improvement of 20\%. We also applied the Kalman detector to existing data from FRB 20201124A and detected two new repeat bursts that were previously missed. Furthermore, we suggest a practical implementation for real-time surveys by employing a low significance soft-trigger from initial flux integration-based detection algorithms. The Kalman detector has the potential to significantly enhance FRB detection capabilities and enable new insights into the spectral properties of these enigmatic astrophysical phenomena.
\end{abstract}


\section{Introduction}
Fast radio bursts (FRBs) are highly intriguing astrophysical phenomenon consisting of radio transients with timescales spanning from hundreds of $\upmu$s to a few ms that have been the focus of significant scientific development over the last decade \citep{Lorimer:2007, Bailes:2022}. These energetic bursts (with isotropic energies in the range $10^{35}$--$10^{42}$ erg) are characterized by large observed dispersion measures (DMs), indicating an extragalactic origin \citep{Thornton:2013, Caleb:2021}. In fact, several FRB sources have been linked to their host galaxies at cosmological distances \citep{Chatterjee:2017, Bannister:2019, Ravi:2019_localized}. A crucial aspect that differentiates the known sample of FRBs is their repeatability; some sources emit multiple bursts, while others seem to produce only single events \citep{Spitler:2016}. Although only a small fraction ($\sim$\,3\%) of FRBs have been observed to emit repeat bursts to date \citep{CHIME:2023}, periodic modulation in burst activity has been confirmed in two sources \citep{CHIME:2020_periodicity, Rajwade:2020}. Radio telescopes with wide coverage areas, such as the Canadian Hydrogen Intensity Mapping Experiment (CHIME) and the Australian Square Kilometer Array Pathfinder (ASKAP), have significantly advanced our observational understanding of FRBs \citep{Shannon:2018, CHIME:2019_400MHz}. CHIME, in particular, has played a pivotal role in this regard, with its capability to make multiple daily detections, leading to the identification of an increasing number of repeating FRBs \citep{CHIME:2021, CHIME:2023}. 

Despite the rapid advancement, our understanding of the underlying nature of FRB emission remains constrained by the current limitations of observational capabilities. This is particularly evident with the low number of well-localized FRBs with associated host galaxy redshifts, as well as the limited number of repeating sources \citep{Petroff:2022}. The low burst yield per source poses challenges in detecting potential periodicity in burst emissions or confidently ruling out its existence. Moreover, recent evidence suggests the existence of an energetic burst population at high redshifts \citep{Ryder:2022}, emphasizing the significance of enhancing detection yield to effectively employ them as cosmological probes \citep{Macquart:2020}. To comprehensively characterize the luminosity distribution of FRBs, it becomes imperative to detect and analyze bursts that lie near or below the detection thresholds of current instruments that may have been missed due to the incompleteness of existing search techniques. Upcoming radio surveys and FRB searches are expected to generate immense volumes of data, potentially reaching several petabytes per day \citep{Hallinan:2019, Vanderlinde:2019}. A primary objective of these efforts is to develop novel methods for detecting radio transients that can outperform existing techniques and investigate the hitherto unexplored parameter space \citep{Petroff:2022}. Achieving maximum completeness in the detection of events buried in the data noise levels is crucial, but equally important is the minimization of false-positive events. Spectral information about the transients is an area where significant progress can be achieved. Presently, the observed patchiness of radio signals in broadband instruments is not effectively utilized in search techniques, providing a promising avenue for improvement. 

Detecting radio signals from FRBs can be a challenging task due to the non-uniform shape of their pulse spectra. This non-uniformity in the radio signal is typically a result of interstellar scintillation, which is induced by wave propagation through an inhomogeneous medium along the line of sight \citep{Rickett:1990, Ravi:2016, Shannon:2018}. Adding to this complexity, most FRB signals are inherently band limited, exhibiting varying degrees of spectral occupancy relative to the bandwidth of the observing telescope \citep{Gourdji:2019, Kumar:2021}. Notably, the FRB signals also exhibit rich and complex frequency-dependent intensity patterns, with repeating FRBs often showcasing pronounced morphological effects \citep{Pleunis:2021}. The non-trivial frequency-dependent intensity leads to a loss of sensitivity when the conventional method of averaging burst signal across the receiver bandwidth is employed.

Current statistical techniques for detecting FRBs rely solely on frequency-averaged spectral intensity, assuming a uniform flux across the frequency band. \citep{CHIME:2018_system, Zackay:2017, Bannister:2019_ascl}. This assumption is not valid for FRB signals, as these techniques were initially developed for detecting broadband emission from radio pulsars \citep{Lorimer:2004}. \citet{Spitler:2012} proposed a detection statistic that characterizes the distribution of signal power in the burst spectrum. In this framework, they introduced the spectral modulation index, representing the normalized standard deviation of spectral intensity as a scoring metric. However, the use case of this technique has been limited to distinguishing narrowband radio frequency interference (RFI) from broadband signals. To improve the detection yield, particularly from repeating FRB sources, it is crucial to develop more sensitive detection statistics that can effectively account for the complex frequency-dependent intensity patterns exhibited by burst signals.

In this paper, we present a novel detection statistic, the Kalman FRB detector, which offers significantly improved sensitivity over simple flux integration. In Section \ref{sec:kalman}, we present a derivation of the detection statistic from basic principles. In Section \ref{sec:application}, we demonstrate the applicability of the Kalman detection score to real FRBs and radio telescopes and quantify the potential sensitivity gains. Finally, in Section \ref{sec:recommendation}, we propose a practical way to integrate the Kalman detector into real-time FRB search systems. 

\begin{figure}
\includegraphics[width=\linewidth]{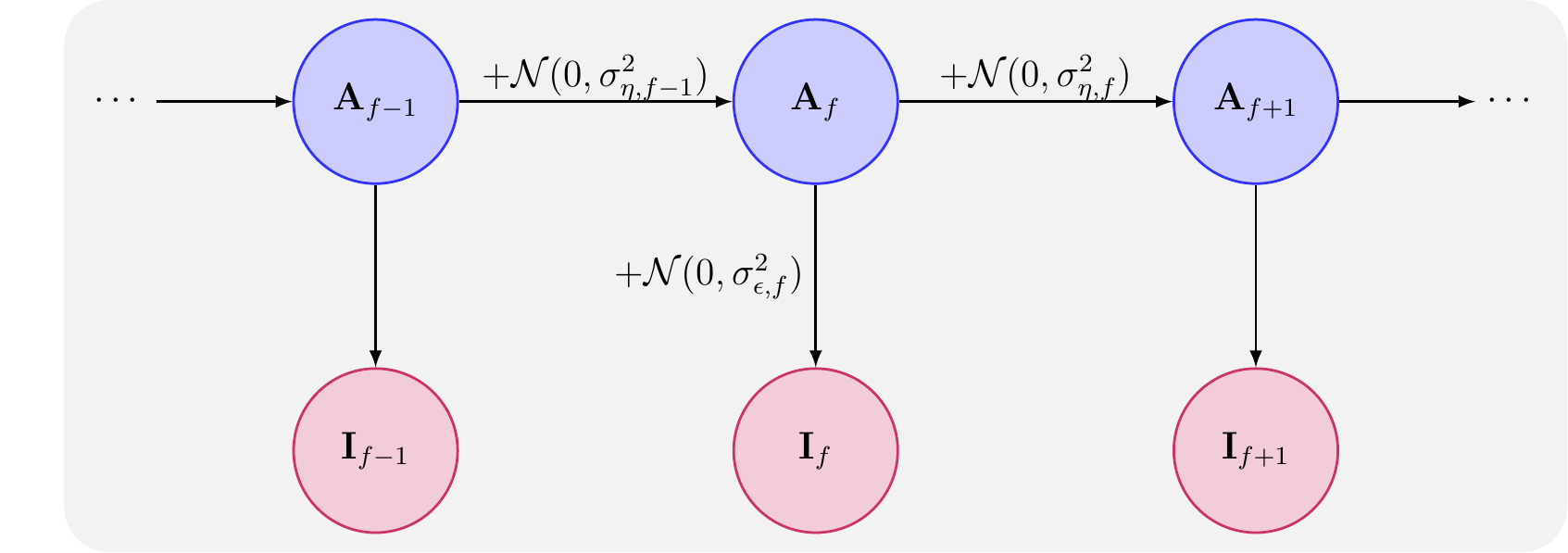}
\caption{Hidden Markov model representation of the dedispersed burst spectrum $I_{f}$, with observed spectra in each frequency channel consisting of the received signal $A_{f}$ and background radio noise.
\label{fig:hmm}}
\end{figure}

\section{The Kalman Detector}\label{sec:kalman}
We model the dedispersed pulse spectral intensity $I_{f}$ as a combination of the received signal and the background noise as
\begin{align}
I_{f} = A_{f} + \epsilon_{f}\,, && \epsilon_{f} \sim \mathcal{N}(0,\sigma_{\epsilon, f}^2)\,,
\end{align}
where $f$ is the channel index in the range 0 to $N-1$ and $N$ is the total number of frequency channels. Here, we use the notation $I(f) = I_{f}$ and $\sigma_{\epsilon} (f) = \sigma_{\epsilon, f}$. The burst signal amplitude as a function of the observed frequency is denoted by $A_{f}$ and $\mathcal{N}(\mu,\sigma^2)$ is the normal distributed noise with expectancy $\mu$ and standard deviation $\sigma$. 
It is important to note that the received signal $A_{f}$ cannot be known in advance. Given our current understanding of FRBs, $A_{f}$ is neither a constant amplitude nor an arbitrarily random function \citep{Pleunis:2021}. This is because $A_{f}$ contains non-trivial amplitude correlations between adjacent frequencies. In order to address this challenge, we propose approximating $A_{f}$ using a Gaussian process or a random walk, defined as 
\begin{align}
A_{f+1} \sim A_{f} + \eta_{f}\,, && \eta_{f} \sim  \mathcal{N}(0,\sigma_{\eta, f}^2)\,.
\end{align} 
Here, $\sigma_{\eta, f}$ governs the correlation length of the signal, where a smaller value implies a larger correlation length in the spectrum. It is worth noting that while the true statistical process could be more complex and dominated by unknown processes within the source, this approximation provides a useful tool for analyzing the relevant parts of the probability distribution function. Additionally, this approach allows us to model the complex correlations in $A_{f}$ in a computationally efficient manner. In the context of the Markov model framework, $A_{f}$ are the hidden (unknown) states, which we aim to estimate based on a series of measurements $I_{f}$. Figure~\ref{fig:hmm} shows a schematic representation of the model.

With our proposed approximation, signal detection can be optimally done using the Neymann-Pearson test \citep{Neyman:1933}. To begin, let us consider a set of measurements denoted by $\mathcal{I}_{f} = \{I_{0},\dots,I_{f}\}$. Our goal is to compute the likelihood ratio $\Lambda$ of hypothesis $\mathcal{H}_1$ (signal present) and $\mathcal{H}_0$ (no signal), which is given as
\begin{align}
\Lambda &= \frac{\Pr[\mathcal{I}_{N-1} ; \mathcal{H}_1]}{\Pr[\mathcal{I}_{N-1} ; \mathcal{H}_0]} = \frac{\Pr[I_{0},\dots,I_{N-1}; \mathcal{H}_1]}{\Pr[I_{0},\dots,I_{N-1}; \mathcal{H}_0]}\,.
\end{align}
Under the null hypothesis $\mathcal{H}_0$, the intensity measurements for each frequency channel are assumed to be independent of one another, given that the radio noise is uncorrelated. In this case, the joint distribution of the observed spectrum is simply the product of the probability distribution of the individual channel observations: 
\begin{align}
\Pr[I_{0},\dots,I_{N-1}; \mathcal{H}_0] = \prod_{f=0}^{N-1}\frac{1}{\sqrt{2\pi\sigma_{\epsilon, f}^2}}\exp{\bigparen{-\frac{\abs{I_{f}}^2}{2\sigma_{\epsilon, f}^2}}}\,.
\end{align}
Now taking the logarithm on both sides, we obtain
\begin{align}
\log\Lambda = \mathcal{L}_{N-1} + \sum_{f=0}^{N-1}{\frac{\abs{I_{f}}^2}{2\sigma_{\epsilon, f}^2}} + \frac{1}{2}\sum_{f=0}^{N-1}{\log(2\pi\sigma_{\epsilon, f}^2)}\,,
\end{align}
where, for convenience, we define the log-likelihood up to the frequency channel $f$ as 
\begin{align}
\mathcal{L}_{f} = \log\bigparen{\Pr[I_{0},\dots,I_{f} ; \mathcal{H}_1]}\,.
\end{align}
While the terms under $\mathcal{H}_0$ in the previous expressions are independent of $A_{f}$, the term under $\mathcal{H}_1$ appears to be intractable. We can overcome this challenge by expressing the joint distribution of the burst spectrum as follows:
\begin{equation}\label{Eq:jointdist}
\Pr[I_{0},\dots,I_{N-1}; \mathcal{H}_1] = \prod_{f=0}^{N-1} \Pr[I_{f} \mid I_{f-1}, \dots, I_{0}; \mathcal{H}_1]\,.
\end{equation}
We adopt the hidden Markov model (HMM) discussed above, where $A_{f}$ follows a Gaussian distribution given the previous state $A_{f-1}$, as does $I_{f}$. Since the probabilities of transitioning from one state to another involve multiplication, the resulting distribution for all $f$ remains Gaussian. Thus, the conditional distribution of having the received signal $A_{f}$ given the channel observations up to $I_{f-1}$ follows a normal distribution, which we express as
\begin{equation}\label{eq:priordef}
\Pr[A_{f} \mid I_{f-1},\dots,I_{0}] \sim \mathcal{N}(E_{f}, V_{f})\,.
\end{equation}
Here, $E_{f}$ is the prior prediction or expected value of the hidden signal state at channel $f$, based on previous channel observations, while $V_{f}$ denotes the associated variance. Likewise, the probability distribution of $I_{f}$ given all previous observations up to state $I_{f-1}$ can be expressed as
\begin{equation}
\Pr[I_{f} \mid I_{0},\dots,I_{f-1}] \sim \mathcal{N}(E_{f}, V_{f} + \sigma_{\epsilon, f}^2)\,.
\end{equation}
The log-likelihood can then be directly calculated as
\begin{align}
\mathcal{L}_{N-1} = \sum_{f=0}^{N-1}\bigparen{-\frac{1}{2}\log(2\pi(V_{f} + \sigma_{\epsilon, f}^2)) - \frac{(I_{f}-E_{f})^2}{2 (V_{f} + \sigma_{\epsilon, f}^2)}}\,.
\end{align}
Now, all we need is to calculate $E_{f}$ and $V_{f}$ for each frequency channel state to compute the log-likelihood. Surprisingly, these can be determined exactly in linear time using the well-known Kalman filter \citep{Kalman:1960, Shumway:2017}. We can derive a recursion relation for $E_{f}$ and $V_{f}$ by leveraging the properties of the HMM (for a detailed derivation, see Appendix~\ref{sec:linear_kalman_deriv})
\begin{align}
E_{f+1} &= E_{f} + \frac{V_{f}}{V_{f} + \sigma_{\epsilon, f}^2}(I_{f} - E_{f})\,, \label{eq:ef_rec}\\
V_{f+1} &= \sigma_{\eta, f}^2 + \frac{V_{f}}{V_{f} + \sigma_{\epsilon, f}^2}\sigma_{\epsilon, f}^2\,. \label{eq:vf_rec}
\end{align}
The next state estimate is simply obtained by adding the predicted value to the measurement residual multiplied by a factor also known as the Kalman gain. Using Equations \ref{eq:ef_rec}--\ref{eq:vf_rec}, we can calculate the detection statistic $\log\Lambda$ for the observed spectrum, which we refer to as the ``Kalman score." Alternatively, we can interpret the recursion relation as follows: At channel $f+1$, we use the best-predicted distribution based on the entire history and update $E_{f}$ and $V_{f}$ by considering the data loss. This updates the $\mathcal{L}_{f}$ in the process, while the distribution of the data predictor is updated using the previous state and $I_{f}$ as independent sources of information. The process of updating $E_{f}$ and $V_{f}$ to capture all the accumulated knowledge from previous states into a single distribution is generally called the Kalman filter. Therefore, the score, although derived from first principles, essentially is the squared residual difference between the data and the Kalman filter prediction based on the past data. The Kalman detector effectively marginalizes over all possible intensity curves, weighting them by the probability distribution of random walks for a given decorrelation bandwidth. 

\subsection{Binary tree algorithm}
Searching for FRBs or other radio transients in a large survey is a challenging and time-consuming task, mainly due to the need to correct for pulse dispersion caused by the interstellar medium. It requires a blind search by iterating over a range of DM trial values, which can be computationally expensive \citep{Cordes:2003}. However, by integrating the Kalman score calculation into the dispersion correction stage, we can significantly improve the efficiency of the search process. This can be achieved by using the Fast Dispersion Measure Transform (FDMT), a dynamic programming algorithm that efficiently performs dedispersion using a binary tree approach \citep{Zackay:2017}. Combining the FDMT algorithm with the Kalman score involves modifying the addition rule in the FDMT iteration to a Kalman marginalization combination. However, this also requires creating a more complex state representation that can describe the Kalman response for the individual spectral segments. With this approach, we can generate sufficient statistics for each spectral segment, allowing for further combination with other segments to enhance detection power. 

To efficiently calculate the Kalman score of the entire signal spectrum, a bottom-up approach can be used, starting from the bottom level of the tree. The signal spectrum is initially partitioned into pairs of frequency channels, and their Kalman state is calculated. Next, the Kalman state of sets of four-channel blocks is obtained by adding the results of the previous step. This process is repeated recursively, with each iteration involving the addition of the results from the previous step until the Kalman state of the whole spectrum is obtained. A detailed description of the new Kalman state representation, their initialization, and an addition rule is given in Appendix~\ref{ap:TreeKalman}.

\subsection{Determination of significance}
\begin{figure}
\centering
\includegraphics[width=\columnwidth]{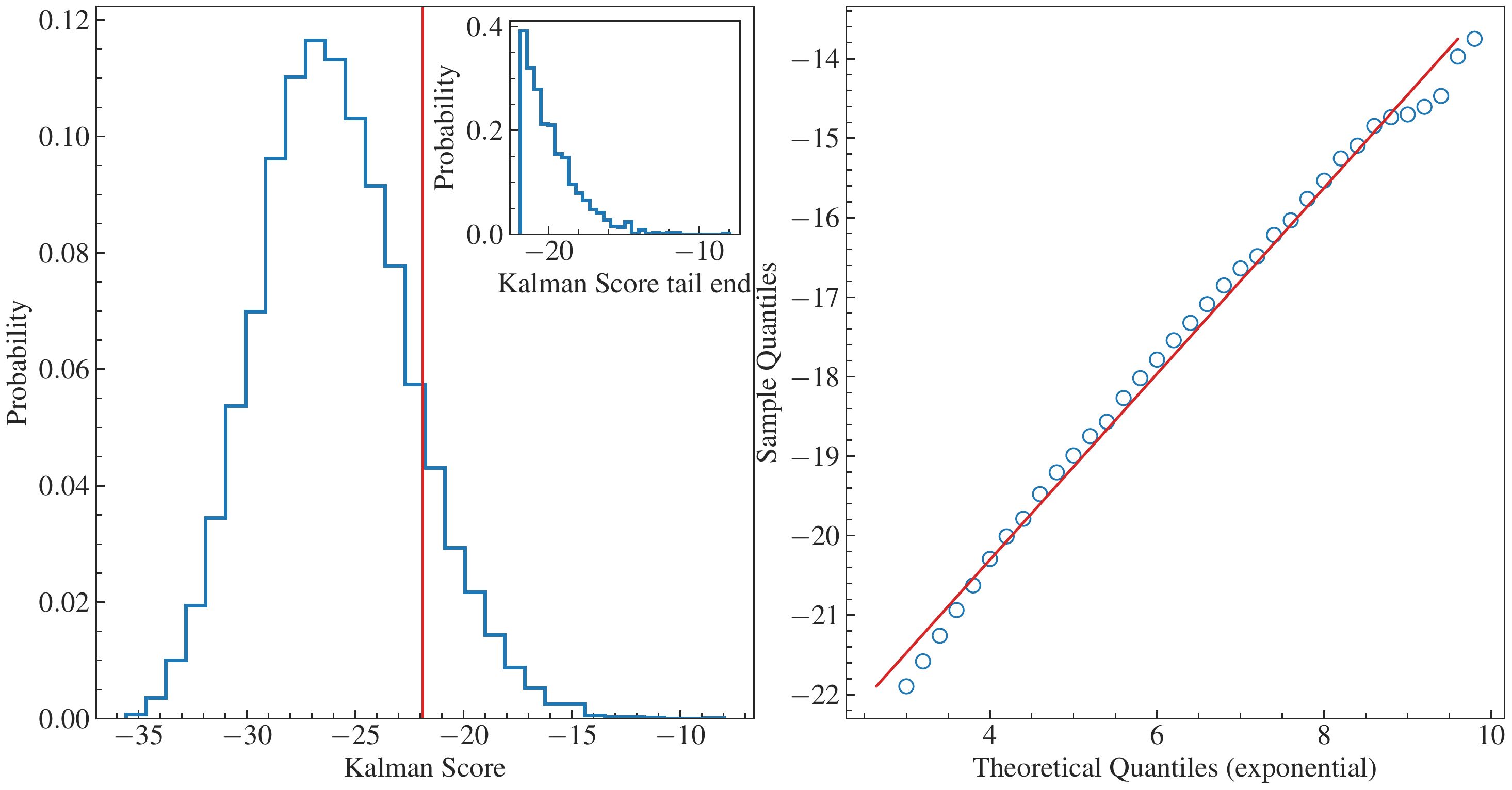} 
\caption{Kalman score distribution and quantile-quantile fit to its tail using a scaled exponential function. The tail of the Kalman distribution is determined using a threshold criterion of 87.5th percentile for the function $y = 2^{-x}$.}
\label{fig:distribution}
\end{figure}

The significance of a real event can be determined by comparing it to the distribution of Kalman scores obtained from the background Gaussian noise. While it might be possible to determine the exact distribution of Kalman scores, in practice, we use Monte Carlo simulation to generate a large number of mean-subtracted Gaussian noise spectra, usually on the order of $\approx 10^4$. This approach allows us to approximate the tail of the distribution by fitting a scaled exponential function of the form $y = 2^{-x}$, which can be easily extrapolated. To accomplish this, we perform a quantile-quantile fit to the distribution of Kalman scores, employing a threshold criterion set at the 87.5th percentile. This provides a good approximation of the tail, allowing us to accurately estimate the probability of observing a false-positive detection at a given significance threshold. When encountering an extreme value of the distribution, we extrapolate the exponential tail to assess its significance. In Figure~\ref{fig:distribution}, we present the distribution of Kalman scores obtained from the Monte Carlo simulations for spectra consisting of 336 channels, along with the corresponding tail fit.

The Kalman detector is effective in handling spectra with unique standard deviation (std) patterns for each channel. However, in such cases, the significance of a detection determined through Monte Carlo simulation may vary depending on the specific std pattern. This would require performing simulations multiple times, which could be computationally expensive. An alternative to the Monte Carlo simulation approach for determining significance is to use a pre-computed std pattern. While this approximation may not be as precise as the simulation method, our tests have shown that it is accurate enough for real-time use. 

In addition, the Kalman score is independent of any constant bias in the spectral data. This statistical independence implies that it can be used in conjunction with the power integration statistic, and the sensitivity of both can be combined while quantifying each contribution separately. However, since the signal and noise components of the two independent sources are not additive, the signal-to-noise ratio (S/N) values cannot be simply added together. Instead, assuming that the noise distributions of both sources are statistically independent and normally distributed, we can combine the tail distribution of the total power statistic $\rm S/N_{P}$ with the Kalman score in the following manner,
\begin{equation}
     \log({\tilde{F}(\rm S/N_{P + K}})) = \log({\tilde{F}(\rm S/N_{P}})) + \log({\tilde{F}(\rm S_{K}})).
\end{equation}
Here, $\tilde{F}$ represents the complementary cumulative distribution function (CCDF) or the survival function, and $\rm S_{K}$ is the significance of the Kalman score obtained using exponential extrapolation. The combined CCDF corresponds to an enhanced S/N compared to considering each factor separately. Finally, we utilize the inverse survival function of the normal distribution to convert this combined CCDF into the corresponding $\rm S/N_{P + K}$.

\subsection{Missing observations}
The Kalman detector offers several advantages over traditional methods for analyzing radio observations, particularly in dealing with missing spectral channels. Radio observations are often affected by RFI, and the severity varies with the telescope location, observing frequencies, and time of observations. Most of the time, these RFI can significantly impact the intensity of frequency channels and result in large chunks of data being unusable. Conventional methods for handling RFI, such as masking or replacing channel values with zero or the mean of unaffected channels, have limitations and can result in multiple discontinuities in the radio spectra of the burst signal.

The Kalman detector provides an effective way to deal with missing spectral observations by using a dynamic model that incorporates information from neighboring frequency channel states. Suppose the observations $I_{f}$ with channel $f=l,\dots,m-1$ are missing for $0 < l < m\leq N-1$. For these channels, we have
\begin{align}
\Pr[A_{f+1} \mid I_{0},\dots,I_{f}] &= \Pr[A_{l} + \sum_{j=l}^{f}\sigma_{\eta, j} \mid I_{0},\dots,I_{l-1}]\nonumber\\
&\sim \mathcal{N}(E_{l}, V_{l} + \sum_{j=l}^{f}\sigma_{\eta, j}^2)\,.
\end{align}
Thus, for the missing channels $f=l,\dots,m-1$, the recursion relation becomes
\begin{align}
E_{f+1} &= E_{f}\,, \\
V_{f+1} &= \sigma_{\eta, f}^2 + V_{f}\,.
\end{align}
The above procedure can be similarly extended for more than one chunk of missing observations.

\begin{figure}
\centering
\includegraphics[width=\columnwidth]{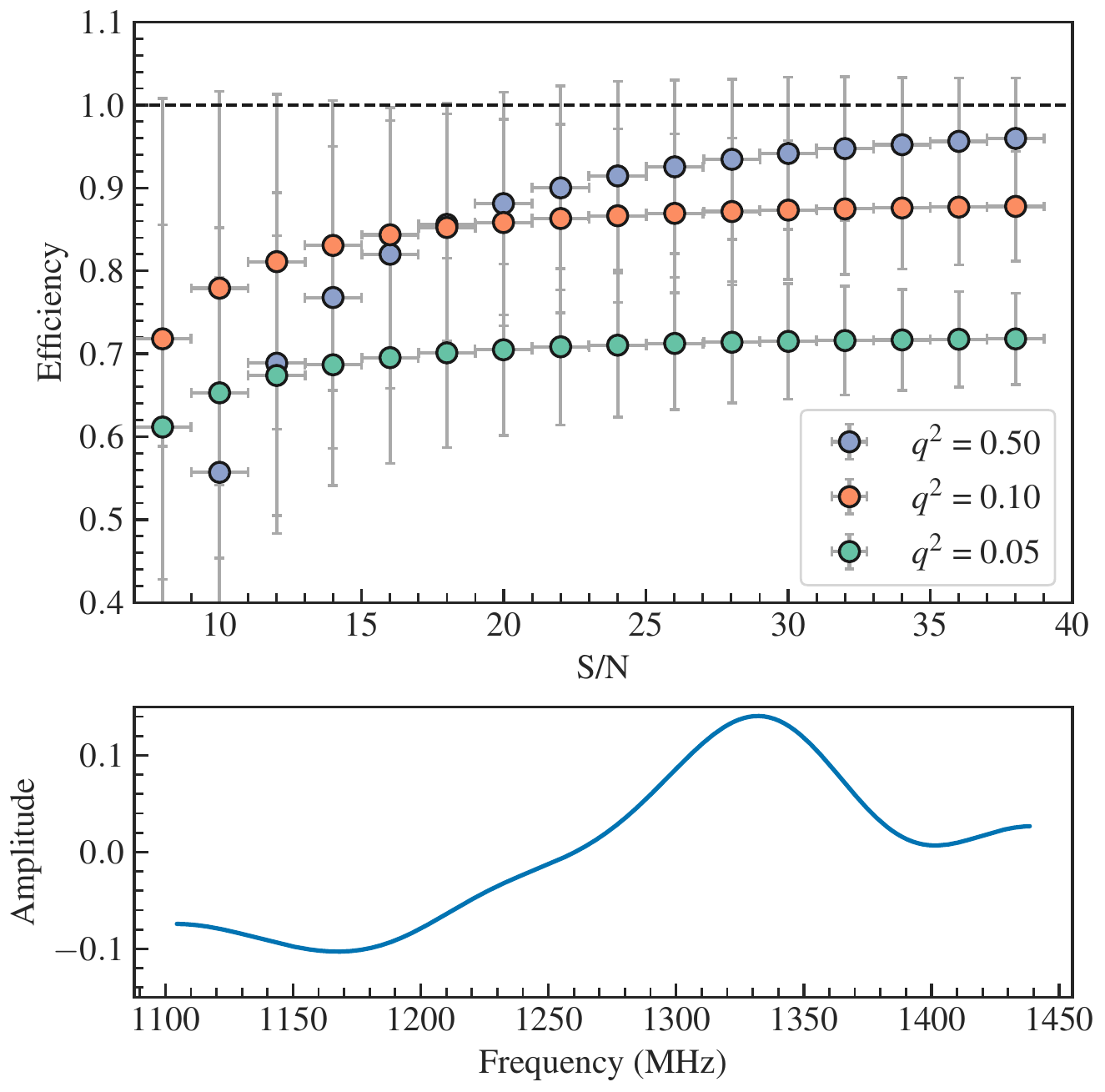} 
\caption{Kalman score efficiency in relation to the optimal matched filter S/N. The top panel illustrates efficiency for a range of $q$ values based on simulated spectra with a mean S/N in the range 6--40. The template FRB spectrum, shown in the bottom panel, serves as the basis for the simulations.}
\label{fig:kalmanEff}
\end{figure}

\subsection{Implementation}
To facilitate the use of the Kalman detector for FRB searches, we provide a user-friendly open-source Python package \soft{kalman\_detector}\footnote{\url{https://github.com/pravirkr/kalman_detector}}. The package includes the implementation of the Kalman detection statistic, along with utility functions for calculating S/N and visualization. We also provide the code implementation for the binary tree algorithm. We have rigorously tested the consistency of the Kalman score obtained using both linear and binary tree approaches for a wide range of input parameters. One important consideration in the implementation is the initialization of the state estimates $E_{f}$ and their variance $V_{f}$. Since the number of frequency channels in FRB spectra data is often large ($> 10^2$), issues of convergence are typically not a concern, and in principle, any initialization would work. In our approach, we opted to initialize the state estimate $E_{f}$ as zero and set the variance $V_{f}$ to the median variance of the spectral noise to reflect bonafide uncertainty in the initial estimates.

\begin{figure*}
\centering
\includegraphics[width=\textwidth]{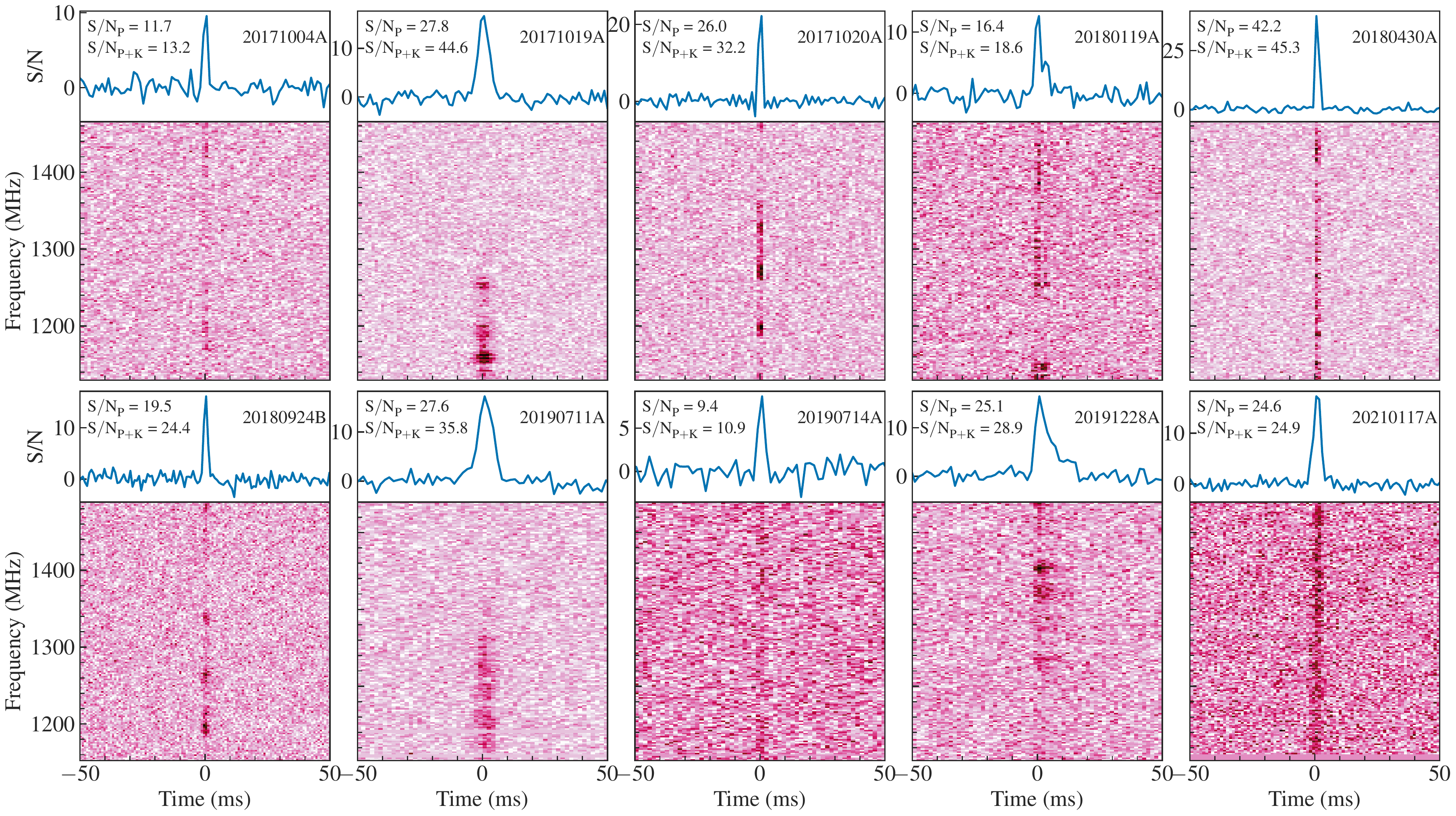} 
\caption{Comparison of S/N for a sample of 10 ASKAP-detected FRBs. All bursts have been dedispersed to their best-reported DM \citep{Shannon:2018}. Each panel displays the dynamic spectrum in the bottom subplot and the frequency-averaged pulse profile in the top subplot. S/N values obtained using both the traditional boxcar filtering (S/N$_{\rm P}$) and the novel Kalman score (S/N$_{\rm P+K}$) are labeled in the top subplot.}
\label{fig:kalmanAskap}
\end{figure*}

Furthermore, we opt to parameterize the process noise $\sigma_{\eta, f}$ in terms of the measurement noise variance $\sigma_{\epsilon, f}$, as follows:
\begin{equation}
q = \frac{\sigma_{\eta, f}^{2}}{\sigma_{\epsilon, f}^{2}}\,.
\end{equation}
In our implementation, we assume that the process noise $\sigma_{\eta}$ is constant across the entire spectrum. Although our theoretical derivations allow for a varying $\sigma_{\eta} (f)$ as a function of frequency, a constant value simplifies Kalman score calculation. To search for the signal over a range of correlation lengths, we calculate the score for a range of $\sigma_{\eta}$ values and report the highest achieved significance. This approach enables us to customize the detection process to better match the characteristics of the signal we are searching for. Furthermore, the user-provided parameter $q$ renders the search process instrument independent. By allowing the user to adjust the value of $q$, the Kalman detector provides flexibility in searching for a variety of transient signals.

Finally, we evaluate the performance of the Kalman filter in comparison to the optimal matched filter. The efficiency of the Kalman score, expressed as its ratio to the matched filter S/N, is illustrated in Figure~\ref{fig:kalmanEff} using a real FRB spectrum. To assess efficiency, we initially derived a smoothed, normalized template from the FRB spectrum. Subsequently, we generate spectra spanning a S/N range of 6--40 by adding noise to the template. For these spectra, we calculate both the Kalman score and the matched filter S/N. Our results demonstrate that, in high S/N scenarios, the Kalman score approaches the optimal S/N, underscoring its status as an optimal test statistic. However, in low S/N situations, measurement noise leads to a reduced score.

\section{Application to the data}\label{sec:application}
We use real FRB data to demonstrate the potential improvement in sensitivity afforded by the Kalman score. We note that the improvement in sensitivity is highly dependent on the signal itself. The score may not provide any improvement for FRBs that do not exhibit considerable frequency structure. Conversely, for FRB signals with prominent frequency structure, the Kalman score can significantly improve sensitivity. As we do not yet have a clear model for FRBs, we chose to test the performance of the Kalman detector on actual FRB signals in order to show its potential for enhancing the detection of these elusive phenomena.

\subsection{ASKAP FRBs}\label{sec:askap_frbs}
We apply the Kalman detection statistic to a set of FRBs previously detected by the ASKAP with a variety of spectral morphologies \citep{Shannon:2018}. We first convolve the frequency-averaged time series of each FRB using a set of normalized boxcar filters to obtain the optimum pulse profile. The best template profile is then used to calculate the integrated S/N$_{\rm P}$. The temporal on-pulse region of the burst profiles is selected using the best-matched template boundary. The rest of the time bins are defined as the off-pulse region. The burst spectrum is then generated by summing the on-pulse region in time of each frequency channel. Similarly, the noise standard deviation associated with the spectrum is calculated using the variance of the off-pulse region. The resulting burst spectra, along with the background noise, are then input to the Kalman detector, which computes the Kalman score based on the spectral shape, and eventually the combined S/N$_{\rm P+K}$. Figure~\ref{fig:kalmanAskap} shows the dynamic spectrum and a comparison of the S/N obtained using the traditional frequency-averaged time series and the novel Kalman score using the spectral shape for a sample of 10 ASKAP-detected FRBs. We find that the Kalman score significantly improves the S/N for some FRBs, particularly those with extreme spectral morphology. For instance, FRB\,20171019A exhibits a significantly band-limited spectrum, and the Kalman score enhances its detection significance by a factor of $\sim$2.5. In contrast, FRB\,20210117A has a featureless spectrum and does not benefit from the application of the Kalman score.

\begin{figure}
\centering
\includegraphics[width=\columnwidth]{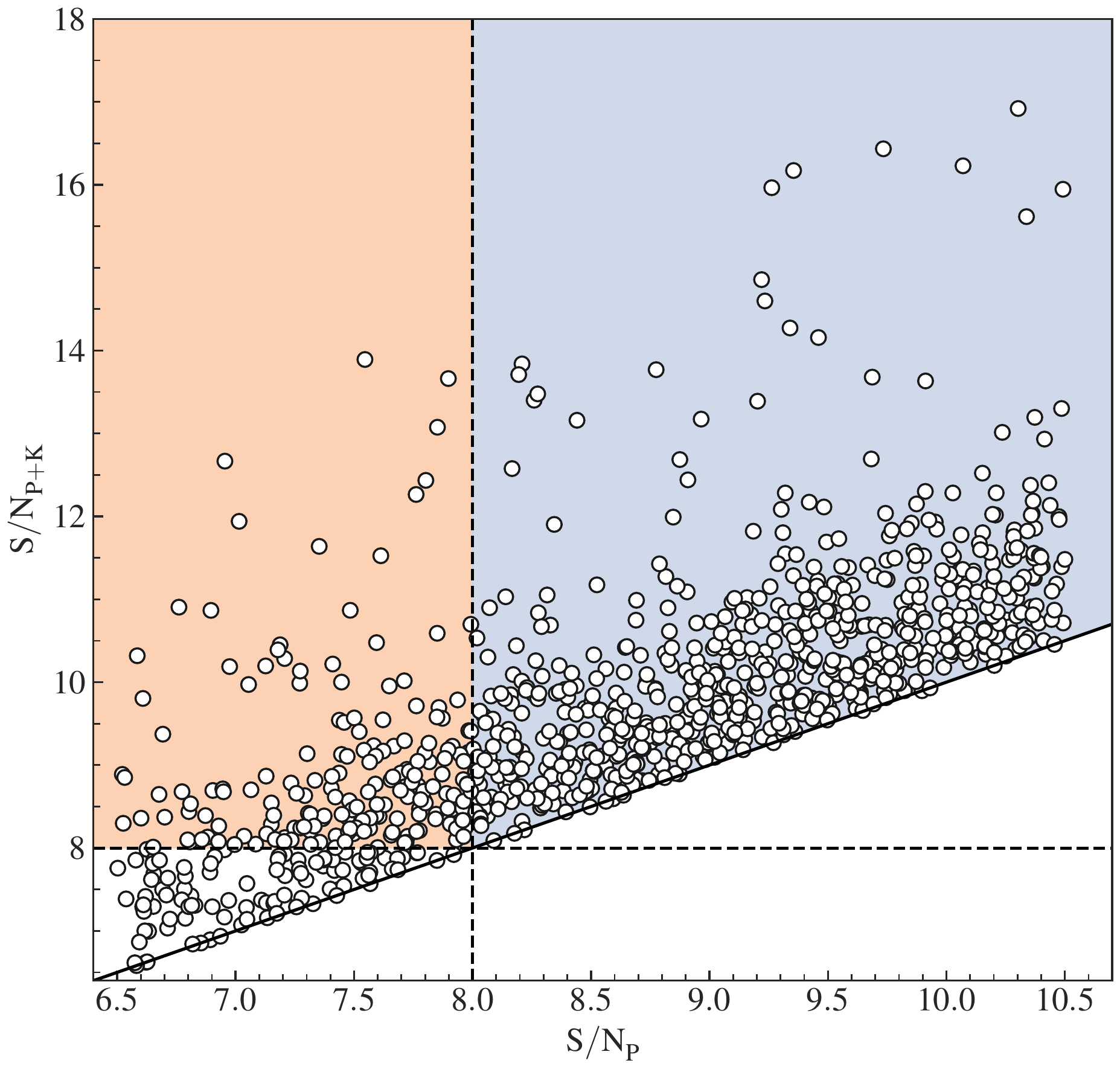} 
\caption{Comparison of S/N obtained using the traditional boxcar filtering and the novel Kalman score method for a sample of 1000 simulated ASKAP-like FRBs with S/N$_{\rm P}$ in the range 6.5--10.5. The S/N$_{\rm P+K}$ values include the contribution from both Kalman score and boxcar filtering. The orange region highlights burst signals that would have gone undetected for the given detection threshold using traditional boxcar filtering alone.}
\label{fig:snr_scatter}
\end{figure}

While the Kalman score has demonstrated its effectiveness in enhancing the S/N of FRBs with non-uniform spectral structures, it is important to assess its utility in comparison to existing techniques, especially for bursts with significance close to the current detection limits. Given the limited number of actual FRB detections, we opt to simulate burst signals to create a well-controlled sample for this analysis. We generated 1000 FRB signals with S/N values in the range of 6.5--10.5 and spectral shapes similar to those of ASKAP-detected FRBs. This is accomplished by adding Gaussian noise to the FRB data. We used the spectral profiles of 32 ASKAP-detected FRBs as templates for these simulations \citep{Bannister:2017, Shannon:2018}. Subsequently, we compared the S/N values obtained using both the traditional frequency-averaged time series and the Kalman score to assess the performance of the Kalman detector. Figure~\ref{fig:snr_scatter} displays the comparison between the S/N$_{\rm P+K}$ and the S/N$_{\rm P}$ for the simulated FRB signals. The results clearly demonstrate that the Kalman score method offers a substantial improvement in S/N for many of the simulated FRBs compared to the traditional boxcar filtering approach. In real-time search systems like the one employed at ASKAP, a minimum S/N threshold is typically used to reduce the number of false-positive candidates. However, this threshold may result in the exclusion of bursts with band-limited signal and S/N$_{\rm P}$ below the threshold, potentially causing them to go undetected. By employing the Kalman score and S/N$_{\rm P+K}$ for detection, the enhanced sensitivity enables the detection of a larger number of such signals. This improved detection capability is illustrated in Figure~\ref{fig:snr_scatter}, considering an example S/N threshold of eight.

\begin{figure}
\centering
\includegraphics[width=\columnwidth]{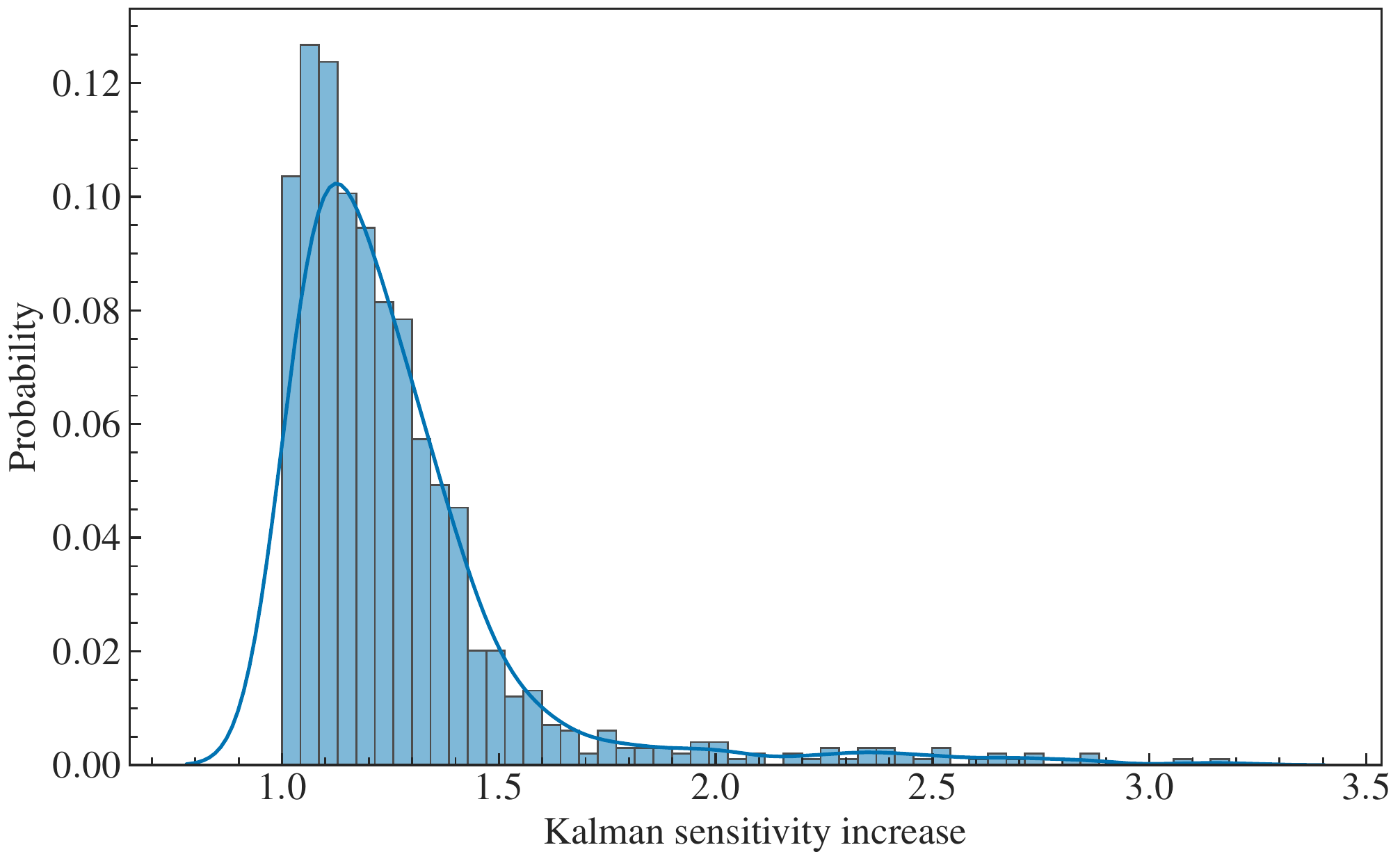} 
\caption{Histogram of sensitivity gain obtained by applying the Kalman score method to a sample of 1000 simulated ASKAP-like FRBs with S/N$_{\rm P}$ in range 6.5--10.5.}
\label{fig:snr_dist}
\end{figure}

We then calculate the total sensitivity gain, defined as the ratio between the significance acquired with the combined boxcar filtering and Kalman score method and the best possible significance obtained from boxcar filtering. The distribution of the sensitivity gain for the simulated events is shown in Figure~\ref{fig:snr_dist}. The histogram demonstrates that the Kalman score technique delivers a considerable sensitivity improvement (up to $\gtrsim$ 200\%) for a majority of the simulated FRBs, with a median gain of 20\%. However, for some FRBs, the Kalman score method does not provide any improvement, as evident by bins with a gain close to one.

\begin{deluxetable*}{ccccccccccc} 
\tabletypesize{\small} 
\tablecolumns{11}
\tablewidth{0pt}
\tablecaption{Burst parameters of the FRB\,20201124A repetitions. Burst properties (except for detection S/N) are measured for the best-fitting sub-band after dedispersing to the DM $= 413$~\DMunits~to be consistent with previous results. For details, refer to \citet{Kumar:2022}. \label{tab:burst_properties}}
\tablehead{ 
\colhead{Burst} & \colhead{TNS event} & \colhead{TOA}  & \colhead{Detection} & \colhead{DM$_\mathrm{S/N}$} & \colhead{$\nu_\mathrm{low}$} & \colhead{$\nu_\mathrm{high}$} & \colhead{Width} & \colhead{Gaussian} & \colhead{Fluence} & \colhead{Peak flux density} \\
\colhead{} & \colhead{FRB} & \colhead{(MJD)} & \colhead{S/N} & \colhead{\DMunits} & \colhead{(MHz)} & \colhead{(MHz)} & \colhead{(ms)} & \colhead{S/N} & \colhead{(Jy ms)} & \colhead{(Jy)}}
\startdata 
K01 & 20210401D & 59305.45546095 & 8.9 & 417 (2) & 796  & -    &  12(2)   & 9.5  & 24(4)  & 2.2(4) \\
K02 & 20210404F & 59308.30919491 & 9.1 & 428 (4) & -    & 1354 &  17(3)   & 8.5  & 25(4)  & 2.3(5)
\enddata
\end{deluxetable*}

\subsection{FRB 20201124A bursts}
Targeted monitoring of the FRB\,20201124A source with the ASKAP was conducted during 2021 April 1--7 \citep{Kumar:2022}. 11 repeat bursts above an S/N of 9 were reported in a total of 16.5 hr of follow-up observations. Given the high burst activity from this source, it presents an ideal data set to search for additional bursts that might have been missed. To identify these potential missed bursts, we reprocessed the ASKAP follow-up data by reducing the S/N threshold and utilizing the Kalman score method. The ASKAP data were originally searched in real-time using the GPU-based detection system \soft{fredda} \citep{Bannister:2019_ascl}. Rather than re-searching the data, we utilized the pre-existing list of candidates generated by \soft{fredda} stored on the disk. First, we employed a friends-of-friends algorithm to cluster together candidates that were detected at different DM trials and time of arrival. We filtered the clustered candidates, keeping DM in the range 150--900\DMunits and S/N$>$8. For each candidate, we then applied the Kalman detector as described in Section~\ref{sec:askap_frbs} to obtain the combined detection significance S/N$_{\rm P+K}$. To reduce the number of false-positive candidates, we additionally masked frequency channels affected by RFI using a modified Z-score threshold of $5.0$. The standard score for each frequency channel was computed using the median absolute deviation of their statistical moments, up to kurtosis.

\begin{figure}
\centering
\includegraphics[width=\columnwidth]{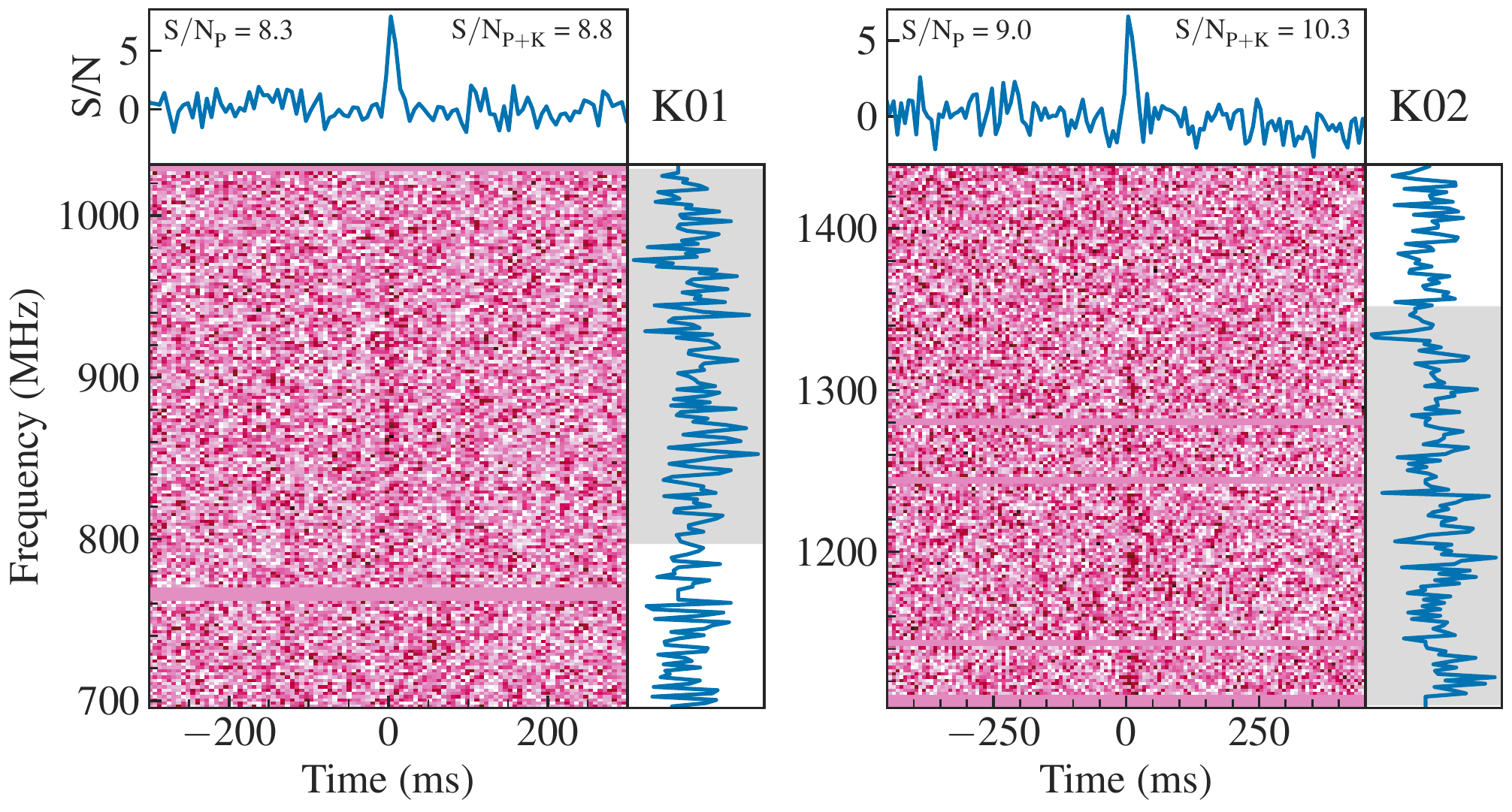} 
\caption{Dynamic spectra of new repeat bursts from FRB\,20201124A, each dedispersed to their best-fitting DM (see Table \ref{tab:burst_properties}). Each panel displays the dynamic spectrum in the bottom subplot (frequency resolution $\sim$2~MHz, time resolution $\sim$6~ms), the time-averaged on-pulse spectrum in the right subplot and the frequency-averaged pulse profile in the top subplot. The gray region in the right subplot indicates the best-fitting sub-band. \label{fig:waterfall_K01_K02}}
\end{figure}

We conducted a visual inspection of all candidates with S/N$_{\rm P+K} > 8$ and identified two previously unreported repeat bursts (K01--K02). In addition, we successfully recovered all bursts that were previously reported, with a notable improvement in significance with a median gain of 80\%. The dynamic spectra of the new repeat bursts are presented in Figure~\ref{fig:waterfall_K01_K02}, and their measured properties are detailed in Table~\ref{tab:burst_properties}. Notably, the \soft{fredda}-reported S/N for burst K02 is well above the detection threshold utilized in the previous search, suggesting that it might have been overlooked in the previous analysis. The detection of burst K01 clearly demonstrates the usefulness of incorporating additional information, such as the Kalman score, in real-time FRB surveys. This further highlights the potential of the Kalman detector to enhance the detection capabilities and improve the identification of previously missed bursts.

\section{Real-time FRB searching}\label{sec:recommendation}
In FRB search surveys, it is critical to detect burst events in real-time to facilitate follow-up with telescopes across various wavelengths and potentially localize the burst progenitor to its host galaxy \citep{Bannister:2019}. However, achieving this real-time detection is challenged by the high computational costs and the considerable number of false-positive candidates that typically arise. Consequently, a minimum S/N threshold (usually within the range of 8--10) is required, depending on the survey parameters of the radio telescope. For optimal FRB detection under the given S/N criterion, one promising approach is the utilization of the Kalman detector, which has shown promising results in the post-processing of FRB data. However, the implementation of the Kalman detector for real-time FRB searches presents several practical challenges. One of the main issues is that the Kalman score is a non-linear function of the data, which means that it requires the FRB pulse profile to be declared ahead of time. This poses a challenge when searching for FRBs exhibiting a wide range of pulse shapes and spectral morphologies, as determining the optimal pulse profile a priori can be difficult.

Another challenge lies in the integration of the Kalman detector into a tree-based algorithm. The FDMT algorithm is a binary tree-based algorithm that is commonly used to probe the DM parameter space and efficiently search for transient events \citep{Zackay:2017}.  Although we provide a tree-compatible version of the Kalman detector in Appendix~\ref{ap:TreeKalman}, with the intention of a potential combination with the FDMT approach, it is important to note that this adaptation demands considerably higher computational resources compared to the standard Kalman detector. As a result, employing a binary tree approach for real-time FRB searches, where computational efficiency is critical, might not be practically viable.

\subsection{Implementation recommendations}
To enhance the detection of FRBs while minimizing computational resources, we present a practical approach for implementing the Kalman detector in real-time FRB searches. We have empirically found that the Kalman score typically does not increase the detection significance by more than a factor of 2 (in principle, this increase is unbounded). This suggests that a high significance threshold can be employed for burst detection. Therefore, we propose the following strategy:

\begin{enumerate}
\item Initiate the search process by applying a standard burst detection algorithm on the input data. Calculate the frequency-averaged detection statistic across a coarse grid of DM and pulse width for each pulse arrival time. For parameter combinations that exceed a threshold of $\approx 3\sigma$ significance, proceed to the next step.
\item Optimize the DM and the pulse width on a finer grid. For candidates that exceed a threshold of $\approx 4\sigma $ equivalent significance (with a false alarm probability of $\sim$10$^{-6}$), move on to the next step.
\item Construct an approximate pulse profile from the frequency-averaged time series. Then, cross-correlate this profile to the individual frequency channels (within the relevant region along the dispersion curve), subtract the mean value, and generate a profile-matched, dedispersed spectrum.
\item Apply the Kalman detector to the spectrum. If the combined significance of the Kalman score and the power-integrated statistic surpasses the survey's detection threshold, classify the pulse candidate as statistically significant.
\end{enumerate}
The thresholds set forth in the aforementioned steps are established based on an ideal search scenario, assuming uncorrelated Gaussian noise. These thresholds are subject to influence from several factors, including the size of the search space (DM, pulse width), available computational resources, and the presence of RFI. Consequently, adapting these thresholds for different instruments might require substantial fine-tuning. With these steps, we avoid the need to declare the FRB pulse profile ahead of time, resulting in a more efficient implementation of the Kalman detector. Furthermore, our approach allows for the integration of the Kalman statistic with conventional burst detection algorithms, enabling its smoother incorporation into real-time FRB searches.

\subsection{VLA/realfast}
The Kalman detector has already been integrated into one of the ongoing FRB search surveys conducted at the Karl G. Jansky Very Large Array (VLA). The survey utilizes a dedicated instrument called \textit{realfast}, which performs commensal, real-time fast transient searching. The \textit{realfast} instrument consists of a specialized compute cluster and a custom fast transient search pipeline \citep{2018ApJS..236....8L}. It operates by sharing fast-sampled (typically $\sim$10\,ms) visibilities generated by the standard VLA correlator with the dedicated \textit{realfast} compute cluster. The compute cluster runs the \soft{rfpipe} software, which performs tasks such as calibration, flagging, dedispersion, and imaging of the fast-sampled visibilities in real time \citep{2017ascl.soft10002L}. By employing a time differencing technique, each generated image becomes highly sensitive to sub-second transients.

Following the initial detection of candidate transients, the \soft{rfpipe} pipeline proceeds to perform additional data quality and statistical tests for each candidate. As part of this candidate analysis, the pipeline calculates the Kalman score for the candidate's spectrum. For each candidate, an estimate of the noise spectrum is generated using a few seconds of visibility data in the vicinity of the candidate. The noise spectrum and the measured spectrum are then utilized to estimate the Kalman score. The resulting image S/N and Kalman score are stored in the candidate management system, enabling the selection of the most significant candidates.

\subsection{Potential caveats}
The flexibility of the Kalman detector makes it more sensitive to narrow-band RFI on the dispersion path, and we note the need for further practical investigation with real data. This extended sensitivity to RFI can also be leveraged as a quality indicator for an RFI-clean pulse environment. Specifically, null Kalman detections with the same DM as a pulse candidate and around the same time may indicate a relatively clean environment, as RFI signals tend to be correlated in both frequency and time.

\section{Summary}
We introduced a novel detection statistic for radio transients with frequency-dependent intensity, which has shown an average increase in sensitivity of 20\% compared to simple flux integration methods for FRBs in our sample. Moreover, we presented a realistic scheme for implementing this statistic in  real-time search systems, allowing for lower detection thresholds and improved detection rates. However, the implementation of this algorithm in real-time searches presents challenges, such as the flexibility required to detect a wide range of pulse shapes and spectral morphologies, as well as the potential for false detections due to narrow-band RFI. Further practical investigations with real data are needed to fully understand these challenges. Nevertheless, the Kalman detector has demonstrated promising results for FRB detection, and its real-time implementation offers a valuable opportunity to probe the unexplored parameter space of the radio sky, particularly in terms of spectral characteristics. Future efforts should focus on developing more efficient and flexible implementations of the Kalman detector that can be seamlessly integrated into existing FRB search algorithms, thereby maximizing its potential for real-time FRB surveys.

\begin{acknowledgments}
We thank Keith Bannister for the helpful discussions. PK and BZ are supported by the Schwartz Reisman Collaborative Science Program, which is supported by the Gerald Schwartz and Heather Reisman Foundation. CJL acknowledges support from the National Science Foundation under grant No.\,2022546. This scientific work uses data obtained from Inyarrimanha Ilgari Bundara / the Murchison Radio-astronomy Observatory. We acknowledge the Wajarri Yamaji People as the Traditional Owners and native title holders of the Observatory site. CSIRO’s ASKAP radio telescope is part of the Australia Telescope National Facility (https://ror.org/05qajvd42). Operation of ASKAP is funded by the Australian Government with support from the National Collaborative Research Infrastructure Strategy. ASKAP uses the resources of the Pawsey Supercomputing Research Centre. Establishment of ASKAP, Inyarrimanha Ilgari Bundara, the CSIRO Murchison Radio-astronomy Observatory and the Pawsey Supercomputing Research Centre are initiatives of the Australian Government, with support from the Government of Western Australia and the Science and Industry Endowment Fund. This research has made use of NASA's Astrophysics Data System Bibliographic Services.

\end{acknowledgments}
\vspace{5mm}
\facilities{ASKAP, VLA}
\software{astropy \citep{Astropy:2013, Astropy:2018},  
          numpy \citep{Harris:2020_numpy},
          matplotlib \citep{Hunter:2007_matplotlib},
          CMASHER for colormaps \citep{Velden:2020_cmasher}
          }

\appendix 

\section{The Continuous Forward Algorithm} \label{sec:linear_kalman_deriv}
The first step in deriving a recursion relation for $E_{f}$ and $V_{f}$ involves noticing the following probability relation,
\begin{align}\label{eq:A1}
\Pr[A_{f} \mid I_{0},\dots,I_{f}] = \frac{\Pr[I_{f} \mid A_{f}]\Pr[A_{f}\mid I_{0},\dots,I_{f-1}]}{\Pr[I_{f} \mid I_{0},\dots,I_{f-1}]}\,.
\end{align}
Here, we used the d-separation property of HMM \citep{Bishop:2006}; given the hidden state, the observation state is independent of all previous states, i.e $I_{f} \perp (I_{0},\dots,I_{f-1}) \mid A_{f}$. Since the above probability relation involves only the multiplication of distributions from the statistical model, which are all Gaussian, $\Pr[A_{f} \mid I_{0},\dots,I_{f}] $ is also Gaussian for all $f$, expressed as
\begin{align}\label{eq:A2}
\Pr[A_{f} \mid I_{0},\dots,I_{f}] &\sim \mathcal{N}(E_{f+1}, V_{f+1} - \sigma_{\eta, f}^2)\,.
\end{align}
After equating the exponential terms of Equations~\ref{eq:A1} and \ref{eq:A2}, we obtained
\begin{align}
\frac{(A_{f} - E_{f+1})^2}{V_{f+1} - \sigma_{\eta, f}^2} &= \frac{(I_{f} - A_{f})^2}{\sigma_{\epsilon, f}^2} + \frac{(A_{f} - E_{f})^2}{V_{f}} - \frac{(I_{f} - E_{f})^2}{V_{f} + \sigma_{\epsilon, f}^2}\\&= \bigparen{\frac{V_{f} + \sigma_{\epsilon, f}^2}{V_{f}\sigma_{\epsilon, f}^2}} \bigbracket{\bigparen{A_{f} - E_{f}} - \frac{V_{f}(I_{f} - E_{f})}{V_{f} + \sigma_{\epsilon, f}^2}}^2\,.
\end{align}
Thus, we have the recursion relation,
\begin{align}
E_{f+1} &= E_{f} + \frac{V_{f}}{V_{f} + \sigma_{\epsilon, f}^2}(I_{f} - E_{f})\,,\\
V_{f+1} &= \sigma_{\eta, f}^2 + \frac{V_{f}}{V_{f} + \sigma_{\epsilon, f}^2}\sigma_{\epsilon, f}^2\,.
\end{align}
Alternatively, the above relation can also be expressed as the weighted average of the state node paths traversed within the HMM model, 
\begin{align}
E_{f+1} &= \bigparen{\frac{1}{V_{f}} + \frac{1}{\sigma_{\epsilon, f}^2}}^{-1}\bigparen{\frac{E_{f}}{V_{f}} + \frac{I_{f}}{\sigma_{\epsilon, f}^2}}\,,\\
V_{f+1} &= \sigma_{\eta, f}^2 + \bigparen{\frac{1}{V_{f}} + \frac{1}{\sigma_{\epsilon, f}^2}}^{-1}\,.
\end{align}

\section{Binary Tree approach: Possible Combination with the FDMT algorithm}\label{ap:TreeKalman}
\begin{figure}
\includegraphics[width=\linewidth]{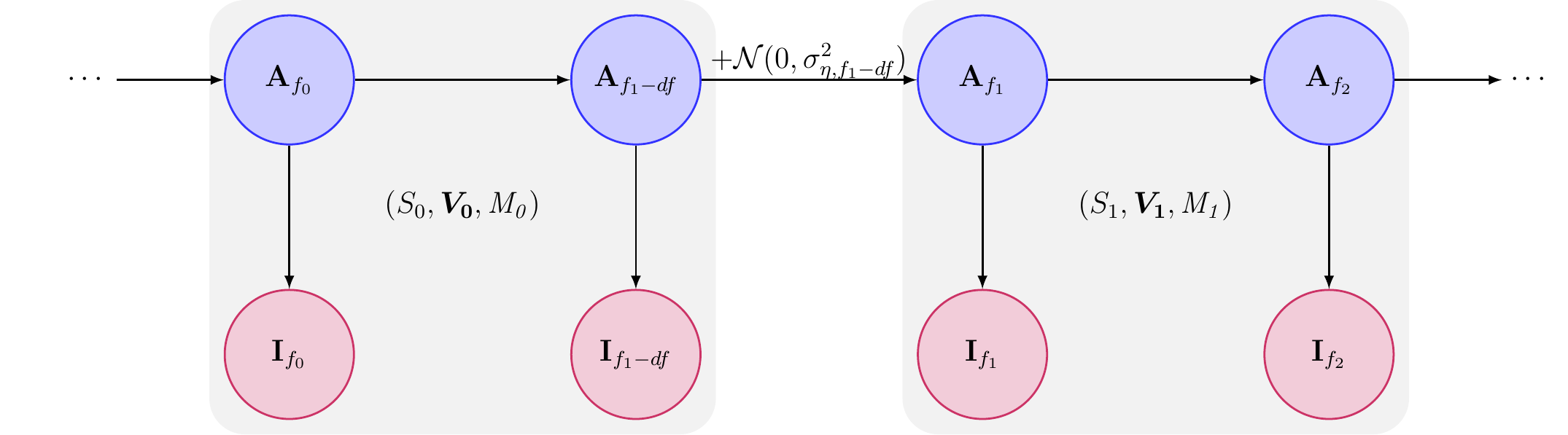}
\caption{Hidden Markov model representation for the binary tree algorithm. The burst spectrum is partitioned into segments, with each segment having a complex Kalman state denoted by the representing parameters $(S, \bm{V}, \mathit{M})$ of a 2D Gaussian distribution.
\label{fig:hmm_binary}}
\end{figure}

The joint probability distribution (given in Equation\,\ref{Eq:jointdist}) for the alternative hypothesis $\mathcal{H}_{1}$ can be marginalized over the received signal model $A_{f}$. The resulting distribution between frequency channels $f_{0}$ and $f_{1}$ can be written as
\begin{align}
\bm{L}_{f_0}^{f_1}(A_{f_0}, A_{f_1}) = \int_{A}{\prod_{f=f_0}^{f_1}{\Pr[A_{f}|A_{f-df}]\Pr[I_{f}|A_{f}]}}\,,
\end{align}
where $A(f_0) = A_{f_0}$, $A(f_1) = A_{f_1}$ and $df$ is the frequency channel bandwidth. Here, we again used the d-separation property of HMM to make variables independent from each other \citep{Bishop:2006}. In other words,
\begin{align}
I_{f} \perp (I_{0},\dots,I_{f-df}) \mid A_{f}\,, && A_{f} \perp (I_{0},\dots,I_{f-df}) \mid A_{f-df}\,.
\end{align}
Using the above formalism, the joint distribution between frequency channels $f_{0}$ and $f_{2}$ (with $f_{0} < f_{1} < f_{2}$) follows the following relation (also see Figure~\ref{fig:hmm_binary})
\begin{align}\label{Eq:recurrence}
\bm{L}_{f_0}^{f_2}(A_{f_0}, A_{f_2}) = \iint_{-\infty}^{\infty}{dA_{f_1-df}\;dA_{f_1}} {\bm{L}_{f_0}^{f_1-df}(A_{f_0}, A_{f_1-df}) \, \bm{L}_{f_1}^{f_2}(A_{f_1}, A_{f_2}) \,  \Pr[A_{f_1}|A_{f_1-df}]}\,.
\end{align}
This recurrence relation leads to the desired linear time calculation of $\bm{L}_{f_{\min}}^{f_{\max}}(A_{f_{\min}},A_{f_{\max}})$. If the probability density functions (PDFs) of $\bm{L}_{f_0}^{f_1-df}(A_{f_0}, A_{f_1-df})$ and $\bm{L}_{f_1}^{f_2}(A_{f_1}, A_{f_2})$ are scaled 2D Gaussians, the resulting $\bm{L}_{f_0}^{f_2}(A_{f_0},A_{f_2})$ also becomes a scaled 2D Gaussian. We thus express our full knowledge of a scaled 2D Gaussian PDF in a set of a scalar, vector, and matrix $(S, \bm{V}, \mathit{M})$. The expression for the resulting distribution can be explicitly written as
\begin{align}
\bm{L}_{f_0}^{f_1}(A_{f_0}, A_{f_1}) = S\exp\left[ - \frac{1}{2} (\bm{A}_{f_0}^{f_1} - \bm{V})^T \mathit{M}(\bm{A}_{f_0}^{f_1} - \bm{V})\right]\,,
\end{align}
where we have denoted $(A_{f_0}, A_{f_1})$ by $\bm{A}_{f_0}^{f_1}$. Here, the scalar $S$ is a normalizing constant that ensures the distribution integrates to unity over the range of possible signal amplitude values. The vector $\bm{V}$ represents the mean vector of a 2D Gaussian distribution, and the matrix $\mathit{M}$ represents the covariance matrix. With these definitions, we can calculate the parameters of the 2D Gaussian distribution resulting from the recursion relation in closed form. The log-likelihood for the spectral segment can be obtained as
\begin{align}
\mathcal{L} (f_{0}, f_{1}) = \log(S) + \log\bigparen{\sqrt\frac{(2\pi)^{2}}{|\mathit{M}|}}\,.
\end{align}

\subsection{Addition rule}\label{ap:addition_rule}
We now describe in detail the computation of the 2D Gaussian distribution parameters of the recursion relation in Equation~\ref{Eq:recurrence}. Let $(S_0, \bm{V_0}, M_0)$ and $(S_1, \bm{V_1}, \mathit{M_1})$ be the representing parameters of the distributions $\bm{L}_{f_0}^{f_1 - df}(A_{f_0},A_{f_1-df})$ and $\bm{L}_{f_1}^{f_2}(A_{f_1},A_{f_2})$, respectively. Furthermore, to ensure consistency between the code implementation and the equations, we access vectors and matrices using the notation $[\;]$, where the first element is denoted by index $0$. In this representation, Equation~\ref{Eq:recurrence} can be written as
\begin{align}\label{Eq:addition_rule_integral}
\bm{L}_{f_0}^{f_2}(A_{f_0}, A_{f_2}) = S_0 S_1 &\iint_{-\infty}^{\infty}dA_{f_1-df}\;dA_{f_1}\times \frac{1}{\sqrt{2\pi\sigma^2_{f_1-df}}} \\ & \nonumber \exp\left[ - \frac{1}{2}(\bm{A}_{f_0}^{f_1 - df} - \bm{V_0})^T \mathit{M_0} (\bm{A}_{f_0}^{f_1 - df} - \bm{V_0}) - \frac{1}{2}(\bm{A}_{f_1}^{f_2} - \bm{V_1})^T \mathit{M_1} (\bm{A}_{f_1}^{f_2} - \bm{V_1}) - \frac{(A_{f_1} - A_{f_1-df})^2}{2\sigma_{\eta, f_1-df}^2}  \right]\,.
\end{align}
The strategy for computing the above integral is to repeatedly use the convolution theorem in the following way,
\begin{align}
\iint_{-\infty}^\infty {dydz G(x,y)K(z-y)H(z,w)} = \int_{-\infty}^\infty {dz H(z,w) \int_{-\infty}^\infty {dy G_x(y)K(z-y)}} = \int_{-\infty}^\infty {dz H(z,w) (G_x \otimes K)[z]}\,,
\end{align}
where $\otimes$ denotes convolution, and $G_x(y) = G(x, y)$. Continuing to evaluate the integral, we obtain
\begin{align}\label{Eq:intConvFormula}
\iint_{-\infty}^\infty {dydz G(x,y)K(z-y)H(z,w)} = (H_w \otimes \overleftarrow{(G_x \otimes K)})[0] = (\mathcal{F}^{-1}(\overline{\mathcal{F}(G_x)}\overline{\mathcal{F}(K)}\mathcal{F}(H_w)))[0]\,,
\end{align}
where the back arrow notation indicates a coordinate change from $z$ to $-z$ and $\mathcal{F}$ represents the Fourier transform. In order to use a closed form to compute the integral, we first need to identify $G_x, K, H_w$. We further simplify the integral as,
\begin{align}
\bm{L}_{f_0}^{f_2}(A_{f_0}, A_{f_2}) &= S_0 S_1 \exp{\left[-\frac{1}{2}(\bm{V_0}^T \mathit{M_0} \bm{V_0}  + \bm{V_1}^T \mathit{M_1} \bm{V_1})\right]} \bm{L}_{\rm int}\,,\\
\bm{L}_{\rm int} &= \iint_{-\infty}^{\infty}dA_{f_1-df}\;dA_{f_1} G_{A_{f_0}} (A_{f_1-df})\, H_{A_{f_2}}(A_{f_1})\, K(A_{f_1} - A_{f_1-df})\,,
\end{align}
where we have identified $G_{A_{f_0}}, H_{A_{f_2}}, K$ as the following
\begin{align}
G_{A_{f_0}} (A_{f_1-df}) &= \exp{\left[-\frac{1}{2}\left( \mathit{M_0}[1,1] A_{f_1-df}^2 +  (2\mathit{M_0}[0,1] A_{f_0}- 2(\bm{V_0}^T\mathit{M_0})[1]) A_{f_1-df} - 2(\bm{V_0}^T\mathit{M_0})[0]A_{f_0} + \mathit{M_0}[0,0]A_{f_0}^2 \right) \right]}\,, \nonumber \\ 
H_{A_{f_2}}(A_{f_1}) &= \exp{\left[-\frac{1}{2}\left( \mathit{M_1}[0,0] A_{f_1}^2  +(2\mathit{M_1}[0,1] A_{f_2}- 2(\bm{V_1}^T\mathit{M_1})[0]) A_{f_1} - 2(\bm{V_1}^T\mathit{M_1})[1]A_{f_2} + \mathit{M_1}[1,1]A_{f_2}^2 \right) \right]}\,, \nonumber \\
&K(A_{f_1} - A_{f_1-df}) = \frac{1}{\sqrt{2\pi\sigma^2_{f_1-df}}}\exp{\left[-\frac{1}{2} \frac{(A_{f_1} - A_{f_1-df})^2}{\sigma_{\eta, f_1-df}^2}\right]}\,.
\end{align}
Now, in order to simplify the Fourier transformation, we express $G_{A_{f_0}}$ in the following form:
\begin{align}
G_{A_{f_0}}(A_{f_1-df}) = \exp{\left[-\frac{1}{2}\left( \frac{(A_{f_1-df} - \beta_0)^2}{\alpha_0} + \gamma_0\right)\right]}\,,
\end{align}
where the terms $\alpha_0$, $\beta_0$, and $\gamma_0$ are given as
\begin{equation}
\begin{aligned}
\alpha_0 &= 1/\mathit{M_0}[1,1]\,, \\ 
\beta_0  &= \alpha_0[(\bm{V_0}^T\mathit{M_0})[1] - \mathit{M_0}[0,1] A_{f_0}]\,, \\
\gamma_0 &= \mathit{M_0}[0,0]A_{f_0}^2 - 2(\bm{V_0}^T\mathit{M_0})[0]A_{f_0} - \frac{\beta_0^2}{\alpha_0}\,.
\end{aligned}
\label{eq: logical 1}
\end{equation}
The Fourier transform of $G_{A_{f_0}}$ is given as
\begin{align}
\mathcal{F}(G_{A_{f_0}})(u) = \sqrt{2\pi \alpha_0}\exp\left[-\frac{1}{2}\left( \alpha_0\,u^2 + 2\,i\beta_0 u + \gamma_0\right)\right]\,.
\end{align}
In the same way, writing $H_w$ as
\begin{align}
H_{A_{f_2}}(A_{f_1}) = \exp{\left[-\frac{1}{2}\left( \frac{(A_{f_1} - \beta_1)^2}{\alpha_1} + \gamma_1\right)\right]}\,,
\end{align}
where the terms $\alpha_1$, $\beta_1$, and $\gamma_1$ are given as
\begin{equation}
\begin{aligned}
\alpha_1 &= 1/\mathit{M_1}[0,0]\,, \\ 
\beta_1  &= \alpha_1 [(\bm{V_1}^T\mathit{M_1})[0] - \mathit{M_1}[0,1]A_{f_2}]\,, \\
\gamma_1 &= \mathit{M_1}[1,1]A_{f_2}^2 - 2(\bm{V_1}^T\mathit{M_1})[1]A_{f_2} - \frac{\beta_1^2}{\alpha_1}\,.
\end{aligned}
\label{eq: logical 2}
\end{equation}
Similarly, the Fourier transform of $H_{A_{f_2}}(A_{f_1})$ is 
\begin{align}
\mathcal{F}(H_{A_{f_2}})(u) = \sqrt{2\pi \alpha_1}\exp\left[-\frac{1}{2}\left( \alpha_1\,u^2 + 2\,i\beta_1 u + \gamma_1\right)\right].
\end{align}
Last, the Fourier transform of $K(A_{f_1} - A_{f_1-1})$ is given as
\begin{align}
\mathcal{F}[K](u) = \exp{\left[-\frac{\sigma_{\eta, f_1-df}^2 u^2}{2}\right]}\,.
\end{align}
Multiplying all of the Fourier transforms together using Equation~\ref{Eq:intConvFormula}, and collecting the coefficients of $u$ we obtain
\begin{align}
\bm{L}_{\rm int} &= \mathcal{F}^{-1}\bigcurly{\sqrt{(2\pi)^2\alpha_0\alpha_1}\exp{\left[-\frac{\alpha_2}{2}u^2 - i \beta_2 u - \frac{\gamma_2}{2}\right]}}[0] = \sqrt{\frac{2\pi\alpha_0\alpha_1}{\alpha_2}}\exp{\left[-\frac{1}{2\alpha_2}(x-\beta_2)^2 - \frac{\gamma_2}{2}\right]}[0]\,,
\end{align}
where the terms $\alpha_2$, $\beta_2$, and $\gamma_2$ are given as,
\begin{equation}
\begin{aligned}
\alpha_2 &= \alpha_0 + \alpha_1 + \sigma_{\eta, f_1-df}^2\,, \\
\beta_2  &= \beta_1 - \beta_0\,, \\
\gamma_2 &= \gamma_0 + \gamma_1\,.
\end{aligned}
\label{eq: logical 3}
\end{equation}
The integral in Equation~\ref{Eq:addition_rule_integral} can be expressed as
\begin{align}
\bm{L}_{f_0}^{f_2}(A_{f_0}, A_{f_2}) = S_0 S_1 &\exp{\left[-\frac{1}{2}(\bm{V_0}^T \mathit{M_0} \bm{V_0}  + \bm{V_1}^T \mathit{M_1} \bm{V_1})\right]} \sqrt{\frac{2\pi\alpha_0\alpha_1}{\alpha_2}}\exp{\left[-\frac{\beta_2^2}{2\alpha_2} - \frac{\gamma_2}{2}\right]}\,.
\end{align}
Now, we want to directly extract the representing parameters of $\bm{L}_{f_0}^{f_2}(A_{f_0}, A_{f_2})$, denoted by ($S_2,\bm{V_2},M_2$). The resulting distribution can be explicitly expressed as follows:
\begin{align}
\bm{L}_{f_0}^{f_2}(A_{f_0}, A_{f_2}) &= S_2\exp\left[ - \frac{1}{2} (\bm{A}_{f_0}^{f_2} - \bm{V_2})^T \mathit{M_2} (\bm{A}_{f_0}^{f_2} - \bm{V_2})\right]\\ &= S_2\exp\bigbracket{- \frac{1}{2} (\bm{V_2}^T\mathit{M_2} \bm{V_2})}\exp\left[ - \frac{1}{2}\bigparen{\bm{A}_{f_0}^{f_2} \cdot \mathit{M_2} \bm{A}_{f_0}^{f_2} - \bm{A}_{f_0}^{f_2} \cdot \mathit{M_2} \bm{V_2} - \bm{V_2} \cdot \mathit{M_2} \bm{A}_{f_0}^{f_2}}\right]\,.
\end{align}
After equating the exponential terms and some algebraic work, we can identify the representing parameters as
\begin{align}
M_2[0,0] &= M_0[0,0] + \frac{M_0[1,0]^{2}}{M_0[1,1]^{2} \sigma_{\eta, f_1-df}^2 + M_0[1,1] + \frac{M_0[1,1]^{2}}{M_1[0,0]}} - \frac{M_0[1,0]^{2}}{M_0[1,1]} \label{eq:M200}\,, \\
M_2[1,1] &= M_1[1,1] + \frac{M_1[1,0]^{2}}{M_1[0,0]^{2} \sigma_{\eta, f_1-df}^2 + M_1[0,0] + \frac{M_1[0,0]^{2}}{M_0[1,1]}} - \frac{M_1[1,0]^{2}}{M_1[0,0]}\,,  \\
M_2[0,1] &= - \frac{M_0[1,0] M_1[1,0]}{M_0[1,1] M_1[0,0] \sigma_{\eta, f_1-df}^2 + M_0[1,1] + M_1[0,0]} \,.
\end{align}
The terms with $A_{f_0}$ and $A_{f_2}$ in the above equation contain the term $M_2 \bm{V_2}$. Collecting these, we obtain
\begin{align}
M_2\bm{V_2}[0] &= M_0[0,0] \bm{V_0}[0]
+ \frac{M_0[1,0]^{2} \bm{V_0}[0]}{M_0[1,1]^{2} \sigma_{\eta, f_1-df}^2 + \frac{M_0[1,1]^{2}}{M_1[0,0]} + M_0[1,1]}
+ \frac{M_0[1,0] M_0[1,1] \bm{V_0}[1]}{M_0[1,1]^{2} \sigma_{\eta, f_1-df}^2 + \frac{M_0[1,1]^{2}}{M_1[0,0]} + M_0[1,1]}\nonumber\\
&- \frac{M_0[1,0]^{2} \bm{V_0}[0]}{M_0[1,1]}
- \frac{M_0[1,0] M_1[1,0] \bm{V_1}[1]}{M_0[1,1] M_1[0,0] \sigma_{\eta, f_1-df}^2 + M_0[1,1] + M_1[0,0]}
- \frac{M_0[1,0] \bm{V_1}[0]}{M_0[1,1] \sigma_{\eta, f_1-df}^2 + \frac{M_0[1,1]}{M_1[0,0]} + 1}\,,\\
M_2\bm{V_2}[1] &= M_1[1,1] \bm{V_1}[1]
+ \frac{M_1[1,0]^{2} \bm{V_1}[1]}{M_1[0,0]^{2} \sigma_{\eta, f_1-df}^2 + \frac{M_1[0,0]^{2}}{M_0[1,1]} + M_1[0,0]}
+ \frac{M_1[0,0] M_1[1,0] \bm{V_1}[0]}{M_1[0,0]^{2} \sigma_{\eta, f_1-df}^2 + \frac{M_1[0,0]^{2}}{M_0[1,1]} + M_1[0,0]}\nonumber\\
&- \frac{M_1[1,0]^{2} \bm{V_1}[1]}{M_1[0,0]}
- \frac{M_0[1,0] M_1[1,0] \bm{V_0}[0]}{M_0[1,1] M_1[0,0] \sigma_{\eta, f_1-df}^2 + M_0[1,1] + M_1[0,0]}
- \frac{M_1[1,0] \bm{V_0}[1]}{M_1[0,0] \sigma_{\eta, f_1-df}^2 + \frac{M_1[0,0]}{M_0[1,1]} + 1}\,.
\end{align}
Taking the inverse of $M_2$ and applying it to both sides of the above equation, we can solve for $\bm{V_2}$. The only remaining parameter to calculate is the normalizing constant $S_2$. After equating the remaining terms, we obtain
\begin{align}
S_2 &= S_0 S_1 \sqrt{2\pi\,C} \exp{\left[-\frac{1}{2}(D - \bm{V_2}^T M_2 \bm{V_2})\right]}\,,
\end{align}
where the scalars $C$ and $D$ are given as
\begin{align}
C &= (M_0[1,1] M_1[0,0] \sigma_{\eta, f_1-df}^2 + M_0[1,1] + M_1[0,0])^{-1}\,,\\
D &= M_0[0,0] \bm{V_0}[0]^{2} +  M_1[1,1] \bm{V_1}[1]^{2}
- \frac{M_0[1,0]^{2} \bm{V_0}[0]^{2}}{M_0[1,1]} - \frac{M_1[1,0]^{2} \bm{V_1}[1]^{2}}{M_1[0,0]}\nonumber\\
&+ \frac{(M_0[1,0]\bm{V_0}[0] + M_0[1,1]\bm{V_0}[1])^{2}}{M_0[1,1]^{2} \sigma_{\eta, f_1-df}^2 + \frac{M_0[1,1]^{2}}{M_1[0,0]} + M_0[1,1]}
- \frac{2 M_0[1,0] M_1[1,0] \bm{V_0}[0] \bm{V_1}[1] - 2 M_0[1,1]M_1[0,0] \bm{V_0}[1] \bm{V_1}[0]}{M_0[1,1] M_1[0,0] \sigma_{\eta, f_1-df}^2 + M_0[1,1] + M_1[0,0]}\nonumber\\
&+ \frac{(M_1[0,0]\bm{V_1}[0] + M_1[1,0]\bm{V_1}[1])^{2}}{M_1[0,0]^{2} \sigma_{\eta, f_1-df}^2 + \frac{M_1[0,0]^{2}}{M_0[1,1]} + M_1[0,0]}
- \frac{2 M_0[1,0]M_1[0,0] \bm{V_0}[0] \bm{V_1}[0] - 2 M_0[1,1]M_1[1,0] \bm{V_0}[1] \bm{V_1}[1]}{M_0[1,1] M_1[0,0] \sigma_{\eta, f_1-df}^2 + M_0[1,1] + M_1[0,0]}\,. \label{eq:D_S2}
\end{align}
For the purpose of reducing the number of computational operations at the expense of potential numerical instability, it is possible to simplify the above equations as follows
\begin{align}
M_2 &= N + CQ\,, \\
M_2\bm{V_2} &= N\bm{U} + CQ\bm{W}\,,\\
D &= \bm{U}^T N\bm{U} + C\bm{W}^T Q \bm{W}\,,
\end{align}
\begin{align}
\bm{Q} =
\begin{bmatrix}
\frac{M_0[1,0]^{2}M_1[0,0]}{M_0[1,1]} & - M_0[1,0] M_1[1,0]  \\[1em]
- M_0[1,0] M_1[1,0] & \frac{M_1[1,0]^{2}M_0[1,1]}{M_1[0,0]}
\end{bmatrix}\,,
\end{align}
\begin{align}
\mathit{N} = 
\begin{bmatrix}
 M_0[0,0] - \frac{M_0[1,0]^{2}}{M_0[1,1]} &  0 \\[1em]
 0 & M_1[1,1] - \frac{M_1[1,0]^{2}}{M_1[0,0]}
\end{bmatrix} \quad
\bm{W} =
\begin{bmatrix}
\bm{V_0}[0] + \frac{M_0[1,1]}{M_0[1,0]}\bm{V_0}[1] \\[1em]
\frac{M_1[0,0]}{M_1[1,0]}\bm{V_1}[0] + \bm{V_1}[1]
\end{bmatrix} \quad
\bm{U} =
\begin{bmatrix}
\bm{V_0}[0] \\[1em]
\bm{V_1}[1]
\end{bmatrix}\,.
\end{align}

\subsection{Initialization}
Before proceeding with the calculation of the Kalman score using the addition rule, it is crucial to establish the appropriate initialization of the $\bm{L}$ array. Specifically, for states with $f \geq 2$, we can write
\begin{align}
\bm{L}_{f}^{f+df}(A_{f},A_{f+df}) &= \Pr[A_{f+df}|A_{f}]  \Pr[I_{f}|A_{f}] \Pr[I_{f+df}|A_{f+df}]\\ \nonumber
&= \frac{1}{\sqrt{(2\pi)^{3}\sigma_{\eta, f}^2 \sigma_{\epsilon, f}^2\sigma_{\epsilon, f+df}^2}} \exp{\left[-\frac{(A_{f+df}-A_{f})^2}{2\sigma_{\eta, f}^2} -\frac{(I_{f}-A_{f})^2}{2\sigma_{\epsilon, f}^2} -\frac{(I_{f+df}-A_{f+df})^2}{2\sigma_{\epsilon, f+df}^2} \right]}\,.
\end{align}
We want to express the above distribution in terms of the 2D Gaussian representing parameters as
\begin{align}
\bm{L}_f^{f+df}(A_{f},A_{f+df}) = S\exp\left[ - \frac{1}{2} (\bm{A}_{f}^{f+df} - \bm{V})^T \mathit{M} (\bm{A}_{f}^{f+df} - \bm{V})\right]\,.
\end{align}
Equating both expressions, we can obtain the representing parameters
\begin{align}
\mathit{M} = 
\begin{bmatrix}
 \frac{1}{\sigma_{\epsilon, f}^2} + \frac{1}{\sigma_{\eta, f}^2} &  - \frac{1}{\sigma_{\eta, f}^2} \\
 - \frac{1}{\sigma_{\eta, f}^2} & \frac{1}{\sigma_{\epsilon, f+df}^2} + \frac{1}{\sigma_{\eta, f}^2}
\end{bmatrix}\,, \quad
\bm{V} = \mathit{M}^{-1}
\begin{bmatrix}
\frac{I_{f}}{\sigma_{\epsilon, f}^2} \\ \frac{I_{f+df}}{\sigma_{\epsilon, f+df}^2}
\end{bmatrix}\,,\\
S = \frac{1}{\sqrt{(2\pi)^{3}\sigma_{\eta, f}^2 \sigma_{\epsilon, f}^2\sigma_{\epsilon, f+df}^2}} \exp\left[ \frac{1}{2} \bm{V}^T\mathit{M}\bm{V} - \frac{I_{f}^2}{2\sigma_{\epsilon, f}^2} - \frac{I_{f+df}^2}{2\sigma_{\epsilon, f+df}^2} \right]\,.
\end{align}
For the channel states $f < 2$, it is essential to incorporate the prior distribution $\Pr[A_{0}] \sim \mathcal{N}(E_{0}, V_{0})$, as defined in Equation~\ref{eq:priordef}. We can explicitly write the distribution as
\begin{align}
\bm{L}_{0}^{1}(A_{0},A_{1}) &= \Pr[A_{1} \mid A_{0}]  \Pr[I_{1} \mid A_{1}] \Pr[A_{1} \mid A_{0}] \Pr[I_{0} \mid A_{0}] \Pr[A_{0}]\\ \nonumber
&= \frac{1}{\sqrt{(2\pi)^{4}\sigma_{\eta, 0}^2 \sigma_{\epsilon, 0}^2\sigma_{\epsilon, 1}^2} V_{0}} \exp{\left[-\frac{(A_{1}-A_{0})^2}{2\sigma_{\eta, 0}^2} -\frac{(I_{0}-A_{0})^2}{2\sigma_{\epsilon, 0}^2} -\frac{(I_{1}-A_{1})^2}{2\sigma_{\epsilon, 1}^2} -\frac{(A_{0}-E_{0})^2}{2V_{0}}\right]}\,.
\end{align}
Expressing above distribution in term of 2D Gaussian formalism, we obtain the representing parameters
\begin{align}
\mathit{M} = 
\begin{bmatrix}
 \frac{1}{\sigma_{\epsilon, 0}^2} + \frac{1}{\sigma_{\eta, 0}^2} + \frac{1}{V_{0}} &  - \frac{1}{\sigma_{\eta, 0}^2} \\
 - \frac{1}{\sigma_{\eta, 0}^2} & \frac{1}{\sigma_{\epsilon, 1}^2} + \frac{1}{\sigma_{\eta, 1}^2}
\end{bmatrix}\,, \quad
\bm{V} = \mathit{M}^{-1}
\begin{bmatrix}
\frac{I_{1}}{\sigma_{\epsilon, 0}^2} + \frac{E_{0}}{V_{0}} \\ \frac{I_{1}}{\sigma_{\epsilon, 1}^2} 
\end{bmatrix}\,,\\
S = \frac{1}{\sqrt{(2\pi)^{4}\sigma_{\eta, 0}^2 \sigma_{\epsilon, 0}^2\sigma_{\epsilon, 1}^2 V_{0}}} \exp\left[ \frac{1}{2} \bm{V}^T\mathit{M}\bm{V} - \frac{I_{0}^2}{2\sigma_{\epsilon, 0}^2} - \frac{I_{1}^2}{2\sigma_{\epsilon, 1}^2}  - \frac{E_{0}^2}{2V_{0}}\right]\,.
\end{align}

\bibliographystyle{aasjournal_pawned}
\bibliography{references.bib}

\end{document}